\documentclass[twocolumn]{revtex4-1}
\usepackage[english]{babel}
\usepackage{amsmath,amsfonts,amsthm} 
\usepackage{graphicx}
\usepackage{xcolor}
\usepackage{physics}
\usepackage{hyperref}
\usepackage{bm}

\keywords{Critical Phenomena, Scaling Hypothesis, Critical Exponents}

\usepackage{amsmath}
\usepackage{epstopdf}
\usepackage{setspace}

\usepackage{xcolor}     
\usepackage{soul}       

\begin{document}

\title{Derivation of the critical point scaling hypothesis using thermodynamics only}

\author{V\'{\i}ctor Romero-Roch\'{i}n}
\affiliation{Instituto de F\'{i}sica, Universidad Nacional Aut\'{o}noma de M\'{e}xico, Apartado Postal 20-364, 
01000 Ciudad de M\'{e}xico, Mexico}

\email{romero@fisica.unam.mx}

\date{\today}

\begin{abstract}
Based on the foundations of thermodynamics and the equilibrium conditions for the coexistence of two phases 
in a magnetic Ising-like system, we show, first, that there is a critical point where the isothermal susceptibility diverges and the specific heat may remain finite, and second, that near the critical point the entropy of the system, and therefore all free energies, do obey scaling. Although we limit ourselves to such a system, we elaborate about the possibilities of finding universality, as well as the precise values of the critical exponents using thermodynamics only. 
\end{abstract}

\maketitle

The scaling hypothesis (SH), introduced by Widom in 1965 \cite{Widom1965}, marked the turning point in the modern description of critical phenomena. It lead to the development of the renormalization group (RG) theory \cite{Kadanoff,Wilson-Kogut} which gave an explanation both to SH and to its deep underlying physics in terms of general concepts such as phenomena at all length scales and symmetry breaking. The trascendental roles of SH and RG cannot be exaggerated, pervading not only the physics and chemistry of phase transitions and condensed matter in general, but also influencing many other fields, from the then emerging field of complex systems to high energy physics. 
 As it is now common knowledge, and expressed in too many articles and monographies, see e.g. Refs. \cite{Fisher-review,Ma,Amit,review-SH}, the scaling hypothesis has remained as such, namely as a {\it hypothesis} that leads to the equalities of the different critical exponents, and that its validation and the actual calculation of the exponents are the success of RG. The purpose of this article is to show that the scaling hypothesis follows directly from the laws of thermodynamics and the equilibrium conditions in a magnetic-like system with a coexistence curve of different thermodynamic states. We show first that the existence of such a curve implies that there is a point, the {\it critical point} of the phase transition, where the thermodynamic properties may or may not be analytic, and where the isothermal susceptibility {\it must} diverge, while the specific heat may remain finite. These results suggest power law dependences of the thermodynamics properties on the natural variables energy and magnetization near the critical point. And as a consequence, the equilibrium conditions at coexistence imply that the entropy function obeys scaling. As we will comment at the end of the text, there may be a way to go further to, first, show that the two critical exponents of the scaling form are not independent of each other, and second, to calculate them without resorting to RG.

To be succinct, we consider the fundamental form of the entropy $s$, per unit of volume, of a very general ``magnetic'' system, with a scalar magnetization $m$ per unit of volume, that can be both positive and negative, and with energy $e$ per unit of volume. That is, we consider the function $s = s(e,m)$ which gives all the thermodynamics of such  a system. Thermodynamics asserts \cite{LLI,Callen,MML} that $s$ is a {\it concave}
 single valued function of $e$ and $m$, with its first derivatives yielding the temperature $T$ and the magnetic field $H$,
\begin{equation}
ds = \beta de - h dm, \label{ds}
\end{equation}
where $\beta = 1/T$ and $h = H/T$,
\begin{equation}
\beta = \left(\frac{\partial s}{\partial e}\right)_{m} \>\>\>\>\>\> h = -  \left(\frac{\partial s}{\partial m}\right)_{e} . \label{betah}
\end{equation}
We consider dimensionless variables \cite{dimensionless}. 
The fact that $s$ is a concave function of $e$ and $m$ follows from the second law and it is expressed through the inequalities \cite{LLI,Callen}
\begin{eqnarray}
- \beta \chi^{-1} &=& \frac{\partial^2 s}{\partial m^2} - \frac{\left(\frac{\partial^2 s}{\partial m\partial e}\right)}{\frac{\partial^2 s}{\partial e^2}} < 0 \nonumber \\
-\beta^2 c_m^{-1} & = & \frac{\partial^2 s}{\partial e^2} < 0, \label{concave}
\end{eqnarray}
which imply $\partial^2 s/\partial m^2 < 0$. In the above inequalities we have already identified the isothermal magnetic susceptibility,
\begin{equation}
\chi = \left(\frac{\partial m}{\partial H}\right)_T \label{chi}
\end{equation}
and the specific heat at constant magnetization,
\begin{equation}
c_m = \left(\frac{\partial e}{\partial T}\right)_m. \label{cm}
\end{equation}
The inequalities and the third law, $\beta > 0$, give rise to the well known stability conditions $\chi > 0$ and $c_m > 0$.

So far, the above expressions are quite general. We now consider properties of a ``magnetic'' system. The main consideration is that $s = s(e,m)$ is an even function of $m$, $s(e,m) = s(e,-m)$. Therefore, the magnetic field $h$ is an odd function of $m$, $h(e,m) = -h(e,-m)$. Hence, if $m = 0$ it follows that $h = 0$. Since there are no restrictions in the energy dependence, and using the third law $\beta > 0$, we find that for constant $m$ the entropy is a monotonic, increasing, concave function of $e$; thus, $\beta$ decreases as $e$ increases for fixed $m$. On the the other hand, for fixed energy $e$, $s$ has a maximum at $m = 0$, then decreases monotonically in a concave fashion as $|m|$ increases. 

Now we analyze the geometrical characteristics that the entropy surface $s = s(e,m)$ should have in order to allow for a two-phase coexistence region. First, for the present system and due to its assumed symmetry, we consider the existence of two phases with magnetizations of opposite signs but same entropy and energy \cite{interaction}, namely, the coexistence of states $(m,e,s)$ with $(-m,e,s)$. From the usual considerations of the coexistence of two thermodynamic states \cite{LLI}, the strong requirement is that their temperature $\beta$ and their magnetic field $h$ are the same. Since $h$ is an odd function of $m$, it must then be true that $h = 0$ for all coexistence states. In addition, we assume that the entropy surface represents {\it stable} thermodynamic states only. Hence, these considerations imply that there exists a {\it void region} on the entropy surface $s = s(e,m)$ whose edge define a {\it coexistence} curve; see the figure for a qualitative rendering of this consideration. This curve is symmetric with respect to $m = 0$ and, aside the point at $m = 0$, the rest of the points on the curve represent two coexisting different phases whose magnetizations $m$ have opposite signs. As we will see below, the introduction of such a curve is so disruptive in an otherwise smooth concave surface, that it forces the point $m=0$ on the curve to be ``critical'', in the sense that $\chi$ must diverge and that the function $s = s(e,m)$ must obey scaling in its neighborhood. The point $m = 0$ on the coexistence curve is identified as the critical point, with energy $e_c$,  entropy $s_c$, temperature $\beta_c$ and critical field $h = 0$. It is very important to emphasize again that the piece of surface {\it inside} the coexistence curve simply does not exist: it is a ``hole'' or a ``cut'' on the entropy surface. It is one of the greatest results of statistical physics that such a type of cuts can be shown to exist in interacting atomic systems in the thermodynamic limit \cite{Yang-Lee}.  

\begin{figure*}[htbp]
\begin{center}
\includegraphics[width=0.5\linewidth]{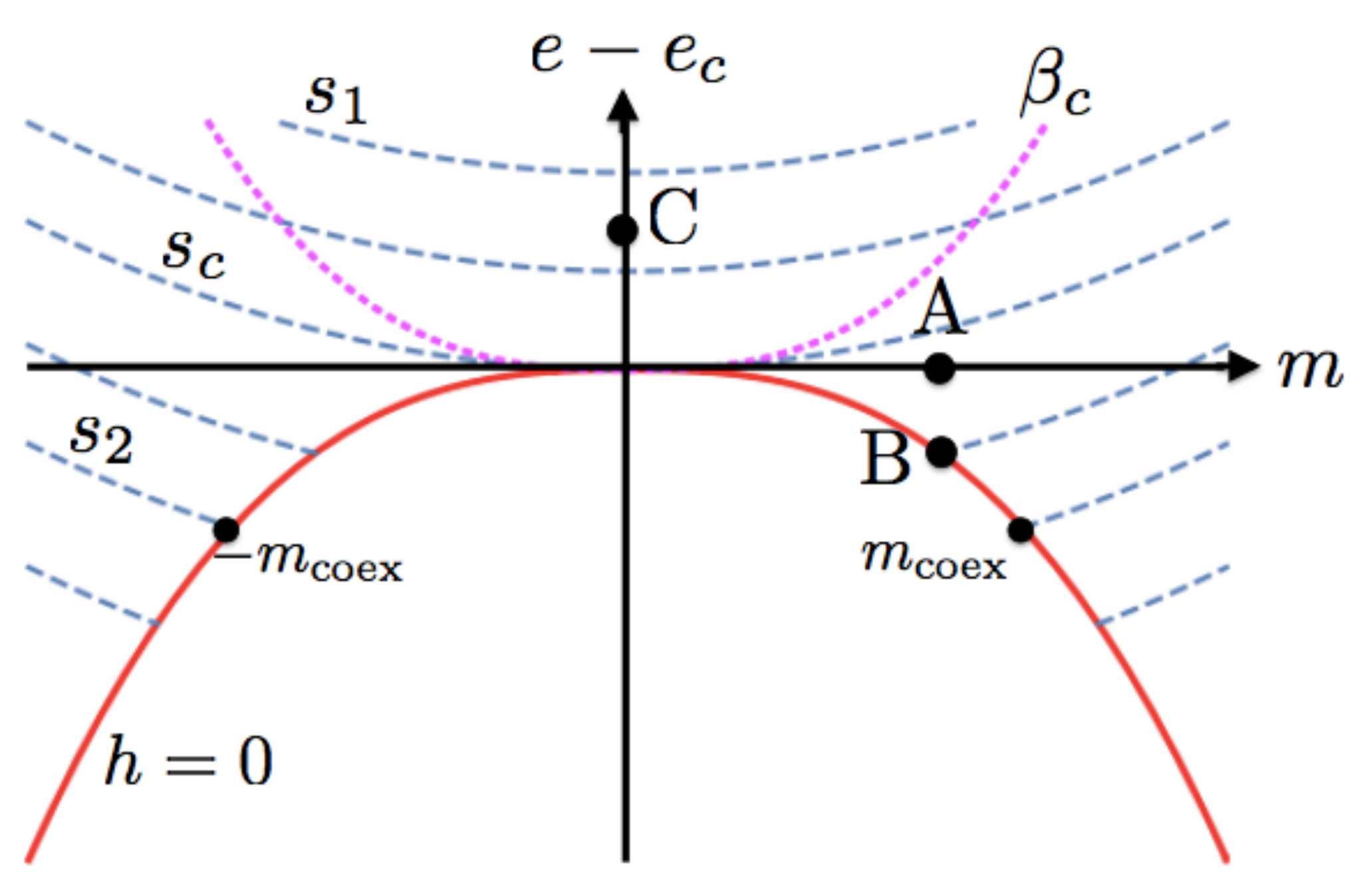}
\end{center}
\caption{(Color online) Qualitative level-curve graph of the entropy $s(e,m)$ in the vicinity of the critical point $e = e_c$ and $m = 0$. The dashed (blue) curves are at constant entropy with $s_2 < s_c < s_1$. The coexistence curve $e-e_c \approx -B (m^2)^\Delta$ is the solid (red) line where the isentropic curves end for $e < e_c$ and the magnetic field vanishes $h = 0$. The labels $-m_{\rm coex}$ and $m_{\rm coex}$ represent two coexisting states with the same entropy, energy, temperature, field $h = 0$, but different magnetizations $\pm m_{\rm coex}$. Within the coexistence curve there is no surface. The point A is at $(e = e_c,m\ne 0)$; point B is at the coexistence curve; and point C at $(e > e_c, m=0$). The dotted (magenta) curve is the critical isothermal $\beta_c$.} \label{fig}
\end{figure*}

Let us now see the consequences  of the coexistence curve. Such a curve can be written quite generally as a relationship between $e$ and $m$,
\begin{equation}
e = e_c + e_{\rm coex}(m^2), \label{ecoex}
\end{equation}
where $e_{\rm coex}(m^2) \le 0$, even in $m$ and vanishing at the critical point. As indicated above $h(e_c + e_{\rm coex}(m^2), m) =0$. Below we will propose an explicit form of such a curve in the neighborhood of the critical point but, first, we can show an important very general result, namely, the vanishing of the inverse of the susceptibility $\chi^{-1}$ at the critical point, see Eqs. (\ref{concave}). For this, let us introduce the vector field $\vec n$ normal to the entropy surface. By considering the vector $(m,e,s)$ oriented in right-hand cartesian axes, a vector normal to the surface at such a point is given by $\vec n(m,e,s) = (h,-\beta,1)$. From the equilibrium conditions one can conclude that at any pair of coexistence points, $s(m,e = e_c + e_{\rm coex}(m^2))$ and $s(-m,e = e_c + e_{\rm coex}(m^2))$, with $m \ne 0$, the corresponding normal vectors are equal. Now consider a point $(m,e,s)$ not on the coexistence curve but near to the critical point $(0,e_c,s_c)$. The normal to such a point can be written as  $\vec n(m,e,s) \approx \vec n_c + (\delta h, -\delta \beta,0)$, where,
\begin{equation}
\delta h = \left(\frac{\partial h}{\partial m}\right)_c \delta m + \left(\frac{\partial h}{\partial e}\right)_c \delta e \label{deltah}
\end{equation}
and 
\begin{equation}
\delta \beta = \left(\frac{\partial \beta}{\partial m}\right)_c \delta m + \left(\frac{\partial \beta}{\partial e}\right)_c \delta e , \label{deltabeta}
\end{equation}
with the subscript $c$ meaning evaluation at the critical point. Consider also the symmetrical point of the above, namely at $(-m,e,s)$; it can be obtained from the previous one by changing $\delta m \to - \delta m$ and leaving $\delta e$ the same. Now we let those two arbitrary points move to the coexistence curve, keeping their symmetry. At coexistence, $\vec n(\delta m, e_c + \delta e,s) = \vec n(-\delta m, e_c +\delta e,s)$, $\delta h = 0$ and $\delta e = e_{\rm coex}(\delta m^2)$. Therefore, from the above expressions, Eqs. (\ref{deltah}) and (\ref{deltabeta}), we conclude that at the critical point,
\begin{equation}
 \left(\frac{\partial h}{\partial m}\right)_c = -\left(\frac{\partial^2 s}{\partial m^2}\right)_c = 0 \label{secm2}
\end{equation}
and
\begin{equation}
\left(\frac{\partial \beta}{\partial m}\right)_c = -\left(\frac{\partial h}{\partial e}\right)_c = \left(\frac{\partial^2 s}{\partial m \partial e}\right)_c = 0 ,\label{secme}
\end{equation}
while there is no restriction on $(\partial \beta/\partial e)_c = (\partial^2 s/\partial e^2)_c$, but to remain negative. By using the concavity conditions, Eqs. (\ref{concave}), and the above critical values, Eqs. (\ref{secm2}) and (\ref{secme}), one finds the severe result that $\chi^{-1} \to 0$ as the critical point is approached, equivalent to assert that $\chi$ diverges there, and that there is no restriction on the (inverse) specific heat $c_m^{-1}$, that is, it may remain finite or not at the critical point. These results are the usual observed behavior at actual phase transitions at the critical point. While the previous derivation makes use of the geometrical properties of the entropy surface, one can visualize this result by noting that the normal vectors at both sides of the coexistence curve must approach the normal at the critical point remaining parallel throughout: intuitively, this can be achieved if the surface at the critical point is flat. Since the gaussian curvature $K = (\partial^2 s/\partial e^2) (\partial^2 s/\partial m^2) - (\partial^2 s/\partial e \partial m)^2$  necessarily vanishes at the critical point, as shown above, the surface at such a point is cylindrical when $(\partial^2 s/\partial e^2)_c \ne 0$ or definitely flat if $(\partial^2 s/\partial e^2)_c = 0$. 

We can now show that the scaling hypothesis is no longer a hypothesis and that it follows from the strong conditions at the coexistence curve. Since the presence of the coexistence curve implies a discontinuity in, at least, one of the variables, there is no reason to expect full analyticity at such a curve.
Thus, while we do not assume analyticity of the entropy function at the coexistence curve and, in particular, at the critical point, we do assume that $s = s(e,m)$ is indeed analytic elsewhere.
In any case, using as the origin the critical point $(e=e_c,m=0)$, we can write the entropy in the form, 
\begin{equation}
s(e,m) = s_c + \beta_c (e - e_c) + s_{\rm sing}(e-e_c,m) ,\label{generals}
\end{equation}
where the function $s_{\rm sing}(e-e_c,m)$ may be singular at $e - e_c = 0$ and $m = 0$, but still $s_{\rm sing}(0,0) = 0$. Since at other points on $s_{\rm sing}(e,m)$ is analytic, we can expand it around the point $e - e_c = 0$ but arbitrary $m\ne 0$, shown as point (A) in the figure,
\begin{equation}
s_{\rm sing}(e-e_c,m) = \sum_{n = 0}^\infty f_n(m^2) \left(e - e_c\right)^n ,\label{ssing}
\end{equation}
where $f_n(m^2)$ are not necessarily analytic at $m = 0$ and, therefore, we cannot make a Taylor expansion around that point. The powerful insight in proposing Eq.(\ref{ssing}) was introduced by Widom in his seminal paper \cite{Widom1965}. The above expression is valid above and at the coexistence curve, such as point (B) in the figure, but not at points such as (C). Thus, we can calculate the isothermal susceptibility $\chi$ and specific heat $c_m$, using Eqs. (\ref{concave}), and the magnetic field $h$, Eq. (\ref{betah}) at any of the valid points. Of particular relevance is their evaluation at the coexistence curve, using Eq. (\ref{ecoex}), where we find,
\begin{eqnarray}
-\beta \chi^{-1} &=& \sum_{n = 0}^\infty \frac{d^2f_n(m^2)}{dm^2} (e_{\rm coex}(m^2))^n -  \nonumber \\&& \frac{\left( \sum_{n = 1}^\infty n \frac{df_n(m^2)}{dm} (e_{\rm coex}(m^2))^{n-1} \right)^2}{ \sum_{n = 2}^\infty n(n-1) f_n(m^2) (e_{\rm coex}(m^2))^{n-2}}. \nonumber \\
-\beta^2 c_m^{-1} &=& \sum_{n = 2}^\infty n(n-1) f_n(m^2) (e_{\rm coex}(m^2))^{n-2} .\label{chicm2}
\end{eqnarray}
The first expression must vanish as $m^2 \to 0$, while the second one may or may not. 
At coexistence, the magnetic field is zero, $h = 0$, yielding the condition,
\begin{equation}
0 = \sum_{n = 0}^\infty \frac{df_n(m^2)}{dm} (e_{\rm coex}(m^2))^n  \label{hatcoex2}
\end{equation}
for all values of $m$.

The previous expressions, Eqs. (\ref{chicm2}) and (\ref{hatcoex2}), and their limits, being functions of $m^2$ only, pose very stringent demands on the form of the coexisting curve and on the functions $f_n(m^2)$. First, one can safely and very generally assume that very near the critical point the coexistence curve is given by
\begin{equation}
e - e_c \approx - B \> (m^2)^\Delta , \label{emcoex}
\end{equation}
where $B > 0$ is a constant and the exponent $\Delta$ is arbitrary. The only reasonable restriction on the exponent is that $\Delta \ge 1$, otherwise it would have a cusp, see the figure. This algebraic dependence and the conditions on the susceptibility at coexistence, indicate that very near $m = 0$ one can asymptotically write 
\begin{equation}
f_n(m^2) \approx A_n (m^2)^{\Gamma_n} , \label{expanm2}
\end{equation}
with $A_n$ constants and the exponents $\Gamma_n$ to be determined. It is important to realize that, even in the asymptotic regime $m \to 0$, due to the possible nonanaliticity of $s_{\rm sing}(e,m)$, in principle one cannot cut the expansion Eq. (\ref{ssing}) at any order $n$ in the sum; that is, the whole sum must vanish in the joint limit $e \to e_c$ first, then $m \to 0$. Hence, expression Eq.(\ref{expanm2}) is the statement that all the functions $f_n(m^2)$ are of equal importance in the expansion and, therefore, that they behave similarly with a (possible nonanalytic) power law behavior near $m =0$. One cannot make any compromise on the coefficients $A_n$, except their contribution to the limiting behaviors of the sums. As it will also be discussed below, there is still room for additional logarithmic terms not considered above.

By substituting the asymptotic expressions given by Eqs. (\ref{emcoex}) and (\ref{expanm2}) into the $h = 0$ condition at coexistence, Eq. (\ref{hatcoex2}), one finds, 
\begin{equation}
0  \approx \sum_{n = 0}^\infty \Gamma_n A_n m^{2\Gamma_n - 1 - 2n \Delta} (-B)^n . \label{h0}
\end{equation}
Since this quantity must be zero for any {\it finite} value of the magnetization $m \ne 0$ near $m = 0$, it can only be so if the exponents at all orders in $n$ are equal, namely, if
\begin{equation}
\Gamma_n - n \Delta = \Gamma_0 \label{cond-scal}
\end{equation}
for all $n = 1, 2, 3, \dots$, and the following sum vanishes,
\begin{equation}
\sum_{n= 0}^\infty (-1)^n \Gamma_n A_n B^n = 0 .
\end{equation}
The condition Eq. (\ref{cond-scal}) for the exponents $\Gamma_n$ yields a scaling form for the entropy function. That is, near the critical point, gathering Eqs. (\ref{ssing}), (\ref{expanm2}) and (\ref{cond-scal}), one finds,
\begin{eqnarray}
s_{\rm sing}(e,m) &\approx& m^{2 \Gamma_0}\sum_{n=0}^\infty A_n \left(\frac{e-e_c}{m^{2\Delta}}\right)^n \nonumber \\
& \equiv & m^{2 \Gamma_0} {\cal F}\left(\frac{e-e_c}{m^{2\Delta}}\right).\label{scaling}
\end{eqnarray}
 The singular part of the entropy is thus expressed in terms of two exponents only, $\Gamma_0 > 1$ and $\Delta \ge 1$, with a scaling function ${\cal F}(x)$. The condition $\Gamma_0 > 1$ follows from the vanishing of $\chi^{-1}$ at criticality, see Eq. (\ref{chicm2}). Obviously ${\cal F}(0) = A_0$, a constant, and, by continuity, $\lim_{x\to \infty}{\cal F}(x)$ must reach $C_0 x^{\Gamma_0/\Delta}$, with $C_0$ a constant. This is a usual argument in dealing with scaling functions\cite{review-SH}, namely, since the form Eq.(\ref{scaling}) is valid everywhere near the critical point, it must also be valid for a point such as (C) in the figure, where $e > e_c$ strictly and $m = 0$, and can depend on $(e-e_c)$ only. Therefore, the limit $m \to 0$ must cancel any dependence on $m$, forcing the mentioned limit. Hence, for $e - e_c >0$, one can expand in powers of $m^2$, yielding,
\begin{equation}
s_{\rm sing}(e,m) \approx (e-e_c)^{\Gamma_0/\Delta}\sum_{k=0}^\infty C_k \left(\frac{m^2}{(e-e_c)^{1/\Delta}}\right)^k ,\label{scaling2}
\end{equation}
with $C_k$ constants that depend on the coefficients $A_n$.

With the forms given by Eqs. (\ref{scaling}) and (\ref{scaling2}) all the scaling results for the thermodynamic quantities near the critical point, including the exponent equalities, follow. See Ref. \cite{review-SH} for a thorough analysis of results following scaling. Since most of the critical properties are typically given in terms of the temperature, one can first find it using Eq. (\ref{betah}); we quote the final results below. But before, we believe it is  instructive to find the critical isotherm curve in the present variables $(e,m)$. For this, let us calculate the temperature near the coexisting curve using Eq. (\ref{scaling}); one finds
\begin{equation}
\beta - \beta_c \approx (m^2)^{\Gamma_0-\Delta} \sum_{n=1}^\infty n A_n \left(\frac{e-e_c}{m^{2\Delta}}\right)^n .\label{betanear}
\end{equation}
By setting $\beta = \beta_c$, $e > e_c$ and $m \ne 0$, the solution is a curve with the same exponent as the coexistence curve, that is $e - e_c \approx D m^{2\Delta}$ but  $D \ne - B$ the factor of the coexistence curve, see Eq. (\ref{emcoex}). See the figure. Using Eq. (\ref{betanear}), it is then a simple exercise to find the usual scaling results in terms of $\Gamma_0$ and $\Delta$, 
\begin{eqnarray}
|m| &\approx& {\cal A} \> (T_c - T)^{1/2(\Gamma_0 - \Delta)} \>\>\>\>h = 0\>\>\>T \le T_c \nonumber \\
h &\approx& {\cal B} \> m^{2 \Gamma_0 - 1} \>\>\>\>T = T_c \nonumber \\
\chi^{-1} &\approx& {\cal C}_{\pm} \> |T-T_c|^{(\Gamma_0-1)/(\Gamma_0-\Delta)} \>\>\>\>h = 0 \nonumber \\
c_m^{-1} &\approx& {\cal D}_{\pm} \> |T-T_c|^{(\Gamma_0-2 \Delta)/(\Gamma_0-\Delta)} \>\>\>\>h = 0 ,\label{exponents}
\end{eqnarray}
where ${\cal A}$, ${\cal B}$, ${\cal C}_{\pm}$ and ${\cal D}_{\pm}$ are constants and the signs $\pm$ indicate $h =0$ and $T > T_c$, and $h = 0$ and $T < T_c$ at coexistence. One reads off the usual critical exponents $\alpha = (\Gamma_0-2 \Delta)/(\Gamma_0-\Delta)$, $\beta = 1/2(\Gamma_0-\Delta)$, $\gamma = (\Gamma_0-1)/(\Gamma_0-\Delta)$ and $\delta = 2 \Gamma_0-1$, obeying the Rushbrooke\cite{Rushbrooke} $\alpha + 2 \beta + \gamma = 2$ and Griffiths\cite{Griffiths} $\beta (1+\delta) = 2-\alpha$ equalities. Two additional comments. If $s_{\rm sing}(e,m)$ is analytic at the critical point, then $\Gamma_0 = 2$ and $\Delta = 1$, the series can be cut at second order and one recovers the usual classical Landau-van der Waals exponents. In general, if $\Gamma_0/\Delta = M$, with $M$ an integer, the series in Eqs. (\ref{scaling}) and (\ref{scaling2}) can be cut at the $M$-th order. For $\Gamma_0/\Delta = 2$
but $s_{\rm sing}(e,m)$ non-analytic, Widom \cite{Widom1965} showed that there could still be a logarithm divergence in the specific heat, as in the two-dimensional Ising model \cite{Onsager}, that we have certainly not considered. But if $\Gamma_0 > 2\Delta$ the logarithmic divergence can be ignored and the specific heat diverges algebraically at the critical point, see Eq. (\ref{chicm2}).

To conclude we first highlight the fact that the scaling form of the entropy near the critical point, as given by Eq. (\ref{scaling}), follows directly from the laws of thermodynamics and its restrictions on the entropy surface. One does not need to introduce it as a {\it hypothesis}. On the other hand, as thermodynamics is an empirical theory that does not explicitly include the dimensionality $d$ of space, it is certainly unable to access the exponents $\eta$ and $\nu$ of the density correlation function \cite{Ma}, and whose relationship to the other exponents is given by the Fisher equality \cite{Fisher-exp} $\gamma =(2-\eta)\nu$  and the hyperscaling Josephson relation \cite{Josephson} $\alpha = 2 - d\nu$. Their elucidation is one of the greatest achievement of RG. However, the present result may open a novel approach to find, in a practical way, the values of the critical exponents, a procedure that we have not been able to materialize. The point being that the exponent $\Delta$ is at our disposal, that is, we can give it any value $\Delta \ge 1$. It is our guess that, again, since the inclusion of the coexistence curve seriously disrupts the otherwise continuous and smooth entropy surface, it may result that such a distortion necessarily requires that the exponent $\Gamma_0$ is a function of $\Delta$, i.e. $\Gamma_0 = \Gamma_0(\Delta)$. Thus, if one were able to find such a relationship, then, by scanning the value of $\Delta$ one could find the value of all the other exponents. The further obvious condition is that the exponent $\Delta$ should be a continuous function of the dimensionality $d$, but at the moment this appears out of context. And finally, the other profound issue of critical phenomena, which adds to the discussion of the thermodynamic origin of the supposed relationship $\Gamma_0 = \Gamma_0(\Delta)$, is the universal character of the critical exponents. The present discussion has been limited to a ``magnetic'' system with its concomitant assumed symmetries. However, as it is well known, the critical exponents of an Ising-magnetic system are the same as those of the critical point of the liquid-vapor transition \cite{Fisher-review,Ma,Amit}. As we will discuss elsewhere \cite{Olascoaga}, one finds that locally the corresponding entropy surface shows the same properties as the present one, once one includes a coexisting curve. Hence, if $\Gamma_0$ is a function of $\Delta$, then, for the same $\Delta$ the exponents will be the same in both physical systems.\\

The author thanks D. Olascoaga for discussions on this matter and support from grant PAPIIT-UNAM IN108620.


\begin{thebibliography}{99}
\bibitem{dimensionless} The dimensionless variables may be, the entropy in Boltzmann constant units, the energy in terms of the coupling constant $J$ of the underlying spins, the magnetization in terms of Bohr magneton and the volume in terms of the crystal constants, say $V = Na$, with $N$ the number of atoms, considered constant throughout.
\bibitem{interaction} In this ensemble, the internal energy $E = Ve$ of an Ising-like magnetic system includes the interaction energy only, namely, $E = - J \sum_{nn} s_i s_j$, with $s_i$ the spin variables and the sum is over nearest-neighbors. So, $e$ is symmetric under the inversion of the magnetization $m = (1/V) \sum_i s_i$.
\end{thebibliography}
\end{document}